\documentclass[aps,notitlepage,nofootinbib]{revtex4-2}

\usepackage{bm,amsmath,amssymb,graphicx}
\usepackage{ amsfonts, bm, epsfig, xcolor, algorithm2e}    

\usepackage[format=plain,labelfont={bf,small},textfont=small,justification=raggedright,singlelinecheck=false]{caption}
\usepackage{subcaption}
\usepackage{tikz}

\begin{document}

\title{Anomalous curvature evolution and geometric regularization of\\ energy focusing in the snapping dynamics of a flexible body}

\author{A. R. Dehadrai} \email{adehadrai@unr.edu}
\author{J. A. Hanna} \email{jhanna@unr.edu}%
\affiliation{ Mechanical Engineering, University of Nevada, Reno, NV 89557-0312, U.S.A. }%

\date{\today}

\begin{abstract}
We examine the focusing of kinetic energy and the amplification of various quantities during the snapping 
motion of the free end of a flexible structure.
This brief but violent event appears to be a regularized finite-time singularity, 
with remarkably large spikes in velocity, acceleration, and tension easily induced by generic initial and boundary conditions.
A numerical scheme for the inextensible string equations is validated against available experimental data for a falling chain and further employed to explore the phenomenon.
We determine that the discretization of the equations, equivalent to the physically discrete problem of a chain, does not provide the regularizing length scale, which in the absence of other physical effects must then arise from the geometry of the problem.
An analytical solution for a geometrically singular limit, a falling perfectly-folded string, accounts surprisingly well for the scalings of several quantities in the numerics, but can only indirectly suggest a behavior for the curvature, one which seems to explain prior experimental data but does not correspond to the evolution of the curvature peak in our system, which instead displays a newly observed anomalously slow scaling.  
 A simple model, incorporating only knowledge of the initial conditions along with the anomalous and singular-limit scalings, 
provides reasonable estimates for the amplifications of relevant quantities.
 This is a first step to predict and harness arbitrarily large energy focusing in structures, with a practical limit set only by length scales present in the discrete mechanical system or the initial conditions.
\end{abstract}

\maketitle

\section{An introduction to snapping}

Energy focusing occurs in a variety of classical settings, among them nonlinear wave phenomena \cite{Peregrine83, DauxoisPeyrard93},  %grimshaw05),
%breathers on lattices , 
sheet crumpling \cite{Witten07}, 
bubble collapse \cite{BarberPutterman92}, %drop detachment \cite{Goldstein93},
 plasma pinches \cite{Bennett34}, and solar flares \cite{LowWolfson88}.
% finite time blowup examples EggersFontelos09
One striking example is encountered often in flexible structure dynamics, where generic initial and boundary conditions generate a rapid and violent ``snapping'' event, in which the free end of a body undergoes a brief whip-like motion associated with large accelerations and a spike in tension.
This phenomenon can be easily observed in everyday life by dropping \cite{CalkinMarch89, Tomaszewski06, GeminardVanel08}, yanking \cite{Bernstein58, Calkin89, Cambou12}, or jiggling \cite{Zak68, Zak70, Belmonte01, Koh99} one end of a chain, or observing a flag flap in the wind \cite{ShelleyZhang11}.
These severe, but far from rare, events have consequences for the fluctuation spectrum of extended bodies with one or more free boundaries, from macromolecular to cosmic scales \cite{Cotta-RamusinoMaddocks10, Bowick17, GoldbergerRothstein06, Armas12}, as well as for the failure of elements in sensing, transport, locomotion, and safety applications.
%towed sonar, in which uncontrollable free end motions are a well-known problem ...  
They are closely related to the pulling taut of cables and tethers in satellites or marine equipment \cite{KaneLevinson77, Zhu99} %Hsu19
and during the deployment of parachute decelerators developed for Mars exploration \cite{OFarrell2019}, as well as to the whipping of whips \cite{GorielyMcMillen02, McMillenGoriely03, Gatti-BonoPerkins02, whipvideo}, in which a tapered cross section contributes an additional amplifying effect on the acceleration. 
%here the focusing is purely due to evolution of the system from initial and boundary conditions

Despite its ubiquity and simplicity of set-up, very little is known about the physics of snapping either from a scientific or practical point of view. 
Just how large are the amplifications of energy density, driving acceleration, and tension?
Experimental and numerical studies, including our own preliminary work, indicate that at least several orders of magnitude are involved.
It has been suggested \cite{Brun16} that this reflects the existence of a regularized singularity and, indeed, the limit of a gravity-driven, perfectly-folded string, whose initial condition involves a geometric singularity in curvature, admits a well-documented analytical solution with finite-time blowup of velocity, acceleration, tension, and tension gradient \cite{Heywood55, CalkinMarch89, OReillyVaradi99, McMillen2005, Virga15, SinghHanna17}. 

What, then, regularizes this singularity?
Physical settings offer many possibilities, such as the bending resistance of cables, the stretching of a bungee or climbing cord as it transitions between inextensible to extensible dynamical regimes \cite{KaganKott96, Strnad97, RealiStefanini96}, the finite mass of a dropped \cite{KaganKott96}, towed, or tethered object, and air drag on a fishing line or its fly \cite{Spolek86, Gatti-BonoPerkins02}. 
Furthermore, experiments on highly flexible bodies feature a discrete multibody system, a chain of links, rather than a truly continuous string. 
%because at practical lab scales, bending regularization is present in any continuum cable/string (the best experimental ``string'' at these scales is often a link chain) 
This discreteness acts similarly to a bending energy, precluding any geometric singularity. 
However, we note that while a nonsingular curvature may focus somewhat during the dynamics, numerical evidence from the recent study by Brun and co-workers \cite{Brun16} based on the unwrapping geometry of Calkin \cite{Calkin89}  suggests that this process is mild. It is not at all obvious that a singularity in curvature should develop for generic smooth initial conditions, although such singularities are possible in some cases \cite{Thess99}. 
%despite assumptions by Brun et al. who look at the curvature presuming singular behavior regularized by bending or discreteness.
  While we will see that curvature evolution is an important ingredient, our present interest is in possible physical singularities in velocity, acceleration, tension, and tension gradient, and their regularization either by physical or numerical effects, or by the geometry of the problem embodied in, for example, the initial conditions.
In the absence of any physical sources of regularization, there remains a numerical length scale associated with the discretization of the problem, which in the present work is entirely akin to a discrete mechanical chain of links.
Is this what regularizes the singularity in numerics? Put another way, for a geometrically regular initial condition, is the physical response singular and, if not, what limits its magnitude?

In the present work, we explore and answer this question through numerical and theoretical approaches to the inextensible string equations, which feature no bending resistance, stretching, or external drag forces.
A finite-difference scheme, featuring a numerical length scale, is validated with available experimental data for the gravity-driven problem, and then employed to examine various additional results. 
We observe the rapid development and disappearance of a small spatiotemporal boundary layer localized at the free end of the body, featuring a very steep tension gradient directly connected with accelerations.
Probing a possible physical singularity presents challenges, as the singularity can only be indirectly inferred as a limit of a physically or numerically regularized system.
We exploit a correspondence between the numerical scheme and a discrete mechanical problem, observe its energy focusing behavior as its length scale is refined, and conclude that this numerical-physical scale is not responsible for the regularization. 
Instead, the small scale is related to the memory of initial geometry and its evolution in time through a previously unobserved and anomalously slow time scaling. 
Concurrently, the velocity, acceleration, and tension evolve predictably through the scalings associated with the ideal geometrically singular problem.
These observations, combined with a simple mechanical model, account for the observed regularization and, with further approximations, allow for reasonable estimates of the remarkably large amplifications.

\section{A model problem, with data}

\begin{figure}[h]
    \centering
    \includegraphics[width=5in]{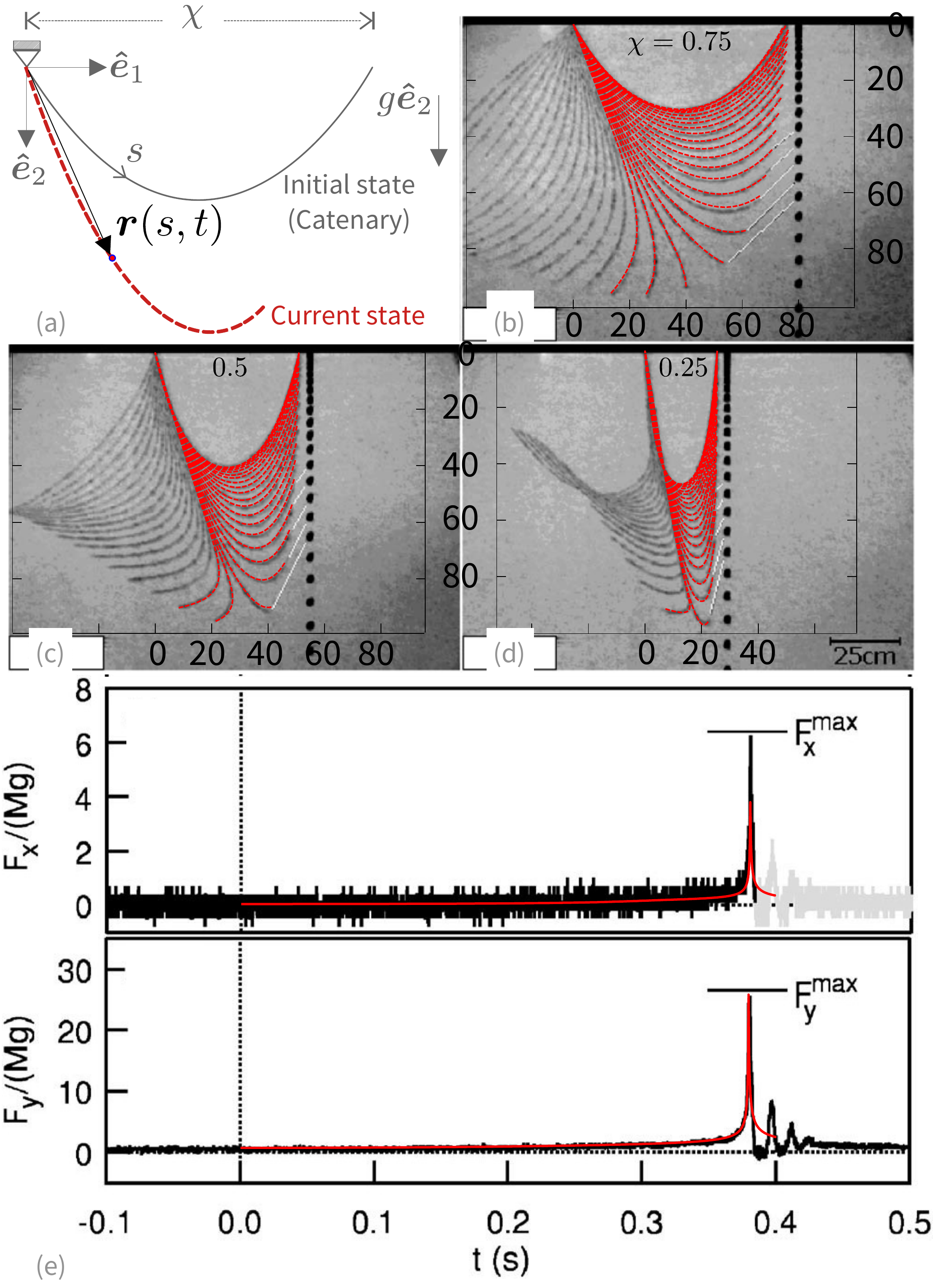}
    \caption{(a) Schematic of the falling string $\bm{r}(s,t)$ of length $\ell$, with initial end-to-end separation $\chi$.  (b)-(d) Computed positions (red) superimposed on experimental images from Tomaszewski and co-workers \cite[Fig.~3]{Tomaszewski06} for $\chi = 0.75\ell, 0.5\ell, 0.25\ell$.   (e) Nondimensional horizontal (upper) and vertical (lower) components of the computed fixed end tension (red) superimposed on experimental data from G{\'{e}}minard and Vanel \cite[Fig.~2]{GeminardVanel08} for $\chi = 0.25\ell$.  In all, $\ell=1.022$ m, $M=\rho \ell = 0.0208$ kg, and $N=2000$. Figures from the references are reproduced with the permission of the American Association of Physics Teachers.}
    \label{validation}
\end{figure}

Snapping may be driven by body forces, distributed drag loading, or momentum input at a boundary. 
We choose to examine the case of a gravity-driven falling catenary with one pinned end, for which detailed experimental data exists to enable comparison.
The equations of motion for an inextensible curve $\bm{r}(s,t)$ parameterized by arc length $s$ and time $t$, with mass density $\rho$, in the presence of gravity $g\bm{\hat{e}}_2$, 
are the balance of momentum
\begin{equation}
d_s \left( \sigma d_s \bm{r} \right) + \rho g \bm{\hat{e}}_2 = \rho  d^2_t  \bm{r} 
\label{eq:eom}
\end{equation}
and the constraint $d_s \bm{r} \cdot d_s \bm{r}  =  1$, which serves to define the tension $\sigma(s,t)$.  Instead of the latter, we employ the tangential projection of the derivative of \eqref{eq:eom} which, after some manipulation involving the original constraint and the permutation of derivatives, is
\begin{equation}
d_s^2 \left( \sigma d_s \bm{r} \right) \cdot d_s \bm{r} = - d_s d_t \bm{r} \cdot d_s d_t \bm{r} \, .
\label{eq:eom2}
\end{equation}
For precedent see \cite{Thess99, Belmonte01, SchagerlBerger02, preston2011, HannaSantangelo12}, and also \cite{EdwardsGoodyear72, Hinch76, GoldsteinLanger95, ShelleyUeda00} for earlier examples with non-inertial dynamics. 
One end of the body is fixed, while the other end is free, with vanishing tension.  For a body of length $\ell$, the boundary conditions are 
\begin{equation}
\bm{r}(0,t) = \bm{0}\, , \quad \sigma(\ell,t) = \bm{0}\,, \quad % \text{\; and \;}  
d_s \sigma(0,t) + \rho g \bm{\hat{e}}_2 
\cdot d_s \bm{r} (0,t)
= 0 \, ,
\label{eq:bcs}
\end{equation}
where the last 
is the tangential projection of \eqref{eq:eom} at $s=0$. 
It will suffice to consider planar dynamics.  
The free end is initially located a distance $\chi$ in the $\bm{\hat{e}}_1$ direction perpendicular to gravity (Fig.\ \ref{validation}(a)).
The initial configuration of the cable
is a catenary, 
$\bm{r}(s,0) = \left[ |a|\sinh^{-1}\left( \tfrac{s-{\ell}/{2} }{|a|}\right) + {\chi}/{2} \right] \bm{\hat{e}}_1
+ \left[ \sqrt{a^2 + ( {\ell}/{2} )^2} - \sqrt{a^2 + (s -{\ell}/{2})^2} \right] \bm{\hat{e}}_2$, 
where the parameter $a$ solves \cite{lockwood1967book}
$\ell = 2a\sinh\left(\frac{\chi}{2a}\right)$ 
and sets the magnitude of the initial maximum curvature through $| \kappa_m(0) | = 1/|a|$.  However, the initial tension is entirely different from that of a catenary with two fixed ends, and is obtained by solving \eqref{eq:eom2} with the initial right hand side set to zero. 
This inextensible model avoids complications arising from an unloading wave triggered by the release of an extensible body's free end to initiate the drop.

We solve the equations with a finite-difference method \cite{preston2011, gatland1994} (Appendix \ref{sec:timeint}). 
%with adaptive time integration, important given the very short time scale and large accelerations during snapping. 
This approach involves no %time-integration, 
physical damping, numerical damping, constraint penalty, or other parameters as would be used in a finite element formulation. % of a (strongly penalized) extensible body. 
More importantly, the formulation is equivalent to one derived from a discrete Lagrangian for a system of $N$ linked pendular masses (Appendix~\ref{sec:timeint}), so that, provided temporal convergence, 
we are actually treating a series of physical, rather than purely numerical, spatially discrete systems approaching the continuum limit. Aside from the number of links $N$, this undamped model has no free parameters. 
We validate this approach (Fig.\ \ref{validation}(b)-(e)) against the available experimental data \cite{Tomaszewski06, GeminardVanel08} for several values of initial end-to-end separation $\chi$. 
The experiments were on chains of length $\ell=1.022$ m, comprising 229 links, with total mass $M=\rho \ell = 0.0208$ kg. 
These are compared with computations performed with $N=2000$; for the deepest catenary in the experiments, this is refined enough to provide tension and velocity values close to the continuum limit. 
%tension very close, velocity is within 3pct, accel 8pct
Results for the geometry, snapping time, and the vertical component of fixed-end tension are in good agreement (neglecting post-snap oscillations associated with the experimental apparatus).  The only significant discrepancy appears to be in the smaller, horizontal component of the end tension, for which we have no explanation.  %almost $30\%$ underestimate. 
These shape results are a better fit than the simulation in \cite{Tomaszewski06} of a 229-link system with \emph{ad hoc} damping employed to fit the end position data only.  
Our values of maximum velocity and, particularly, acceleration differ significantly from this prior simulation, likely due both to the absence of damping as well as being much closer to the continuum limit.
As will be seen shortly, more accurate prediction of acceleration maxima in the continuum limit requires a larger $N$ for this and particularly for deeper catenaries. 

\section{The snapping boundary layer}

\begin{figure}[!h]
    \centering
   \includegraphics[width=6in]{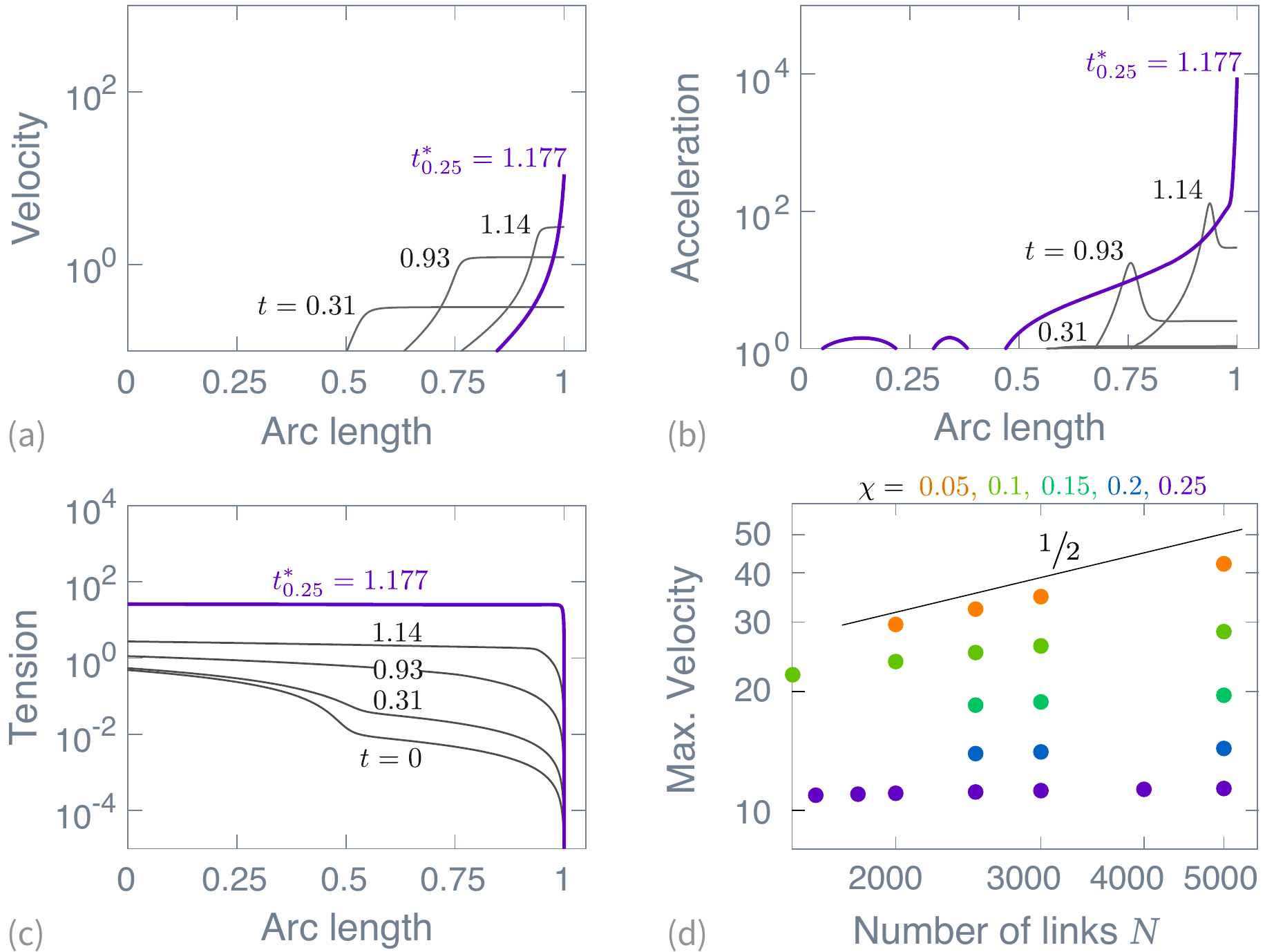}  
   \caption{(a)-(c) Velocity, acceleration, and tension along the body's length at several times leading up to the snapping event (purple), for the $\chi = 0.25$, $N=2000$ case featured in Fig.\ \ref{validation}(d)-(e). 
   (d) The snapping velocity for this and narrower catenaries as a function of $N$, growing more slowly than the $\tfrac{1}{2}$ rate expected of a physical singularity regularized by $N^{-1}$.}
  \label{quantities}
\end{figure}

We hereafter discuss only nondimensional quantities, rescaled with length $\ell$, velocity $\sqrt{g \ell}$, mass density $\rho$, and derived quantities. 
%acceleration $g$, force $\rho g \ell$ and thus time $\sqrt{\ell /g}$ and stress  
Further details in the form of velocity, acceleration, and tension distributions along the moderately deep ($\chi = 0.25$) catenary featured in Fig.\ \ref{validation}(d)-(e) can be seen in Fig.\ \ref{quantities}(a)-(c) for different times, including the snapping event (in purple).
These profiles consist of a smeared propagating front separating two distinct regions.
For most of the time, these regions are separated by a small zone of high curvature reminiscent of the kink in the singular case of a perfectly-folded $(\chi = 0)$ falling string \cite{Heywood55, CalkinMarch89, OReillyVaradi99, McMillen2005, Virga15, SinghHanna17}.
Profiles after snapping retrace qualitatively similar forms, and are not shown.  
The region attached to the tether is much like a static hanging string solution with evolving tension at its lower boundary. The linear tension profile is eventually overshadowed by the rising average tension, presenting the appearance of a plateau as snapping is approached (Fig.\ \ref{quantities}(c)).
 The other region, continuing to the free end, moves primarily tangentially as an almost rigid piece; the extent of energy focusing can be seen from the rising height and shrinking width of this velocity plateau (Fig.\ \ref{quantities}(a)). 
For much of the time before snapping, the peak in acceleration (Fig.\ \ref{quantities}(b)) is located near that of curvature, in the propagating folded portion of the cable.  
We will see later that a late-time shift of this maximum into the free-end region marks a transition to the regularizing whirling motion of this region.
The rising tension conspires with the zero-tension free-end condition to create a large tension gradient in a narrow boundary layer, which shrinks to a very small but finite size before the process halts and reverses over a brief window of time.
It may be recalled or 
gleaned from \eqref{eq:eom} that tangential and normal accelerations respectively correspond to tension gradients and the product of tension and curvature. 
For this moderately-deep catenary, the acceleration reaches order $10^4$ and the snapping process is only about $10^{-3}$ in duration. 

\section{Regularization is not by discreteness}

What regularizes this process, halting and reversing the elbow in tension a short distance from the end?  If the continuous physical system possesses a singularity, it will only be indirectly observable in experiments with discrete chains or numerics involving a finite number of links $N$. 
We assess whether this discretization provides the regularizing length scale by considering how the available energy $P$,  the potential energy of an initial catenary minus that of a vertically hanging string, is focused into a region of length scale $\delta$ that behaves as a rigid rod % (with distributed mass implied), 
rotating with constant angular velocity $\omega$ around a fixed point at $s=1-\delta$.  This yields fields in the boundary layer $v = \omega [s-(1-\delta)]$, $a = \omega^2 [s-(1-\delta)]$, and $\sigma = \tfrac{1}{2}\omega^2\left(\delta^2 - [s-(1-\delta)]^2\right)$, with $P = \tfrac{1}{6}\omega^2\delta^3$, 
  and thus estimates for the peak snapping velocity, acceleration, and tension,% \com{and tension gradient?}: 
\begin{align}
    v_\delta &= \omega \delta = \left( 6 P / \delta \right)^\frac{1}{2}\, ,    \;
    a_\delta = v_\delta^2 / \delta \, ,    \;
    \sigma_\delta = v_\delta^2 / 2 \, , %\;    d_s \sigma_\delta = v_\delta^2  / \delta \, .
    \label{eq:estimates}
\end{align}
neglecting a small adjustment whereby the tension at the tether point is higher due to gravity.
%linear
If the snapping of a falling cable is a singular event, the discrete $N$-link problem would provide the only regularizing length scale in the form $\delta_N= N^{-1}$, and the maximum velocity would scale as $N^\frac{1}{2}$.
Fig.\ \ref{quantities}(d) shows that this is not the case.
Exploring further into more extreme versions of the snapping event for deeper catenaries, we realize that, aside from the analytical solution to the singular case, the maximum values of amplified quantities in the continuum limit can only be approximated by large-$N$ computations.  Yet, while for very deep catenaries the $N$-pendulum problem is only slowly approaching the continuum problem, the increasing velocity maxima clearly fall off from the singular scaling, indicating that there is additional physics or geometry providing the regularization in the continuum limit. 
For moderately deep catenaries, the leveling off of these quantities is already observable at a few thousand links.
Although the data is here shown for initial catenaries as narrow as $\chi = 0.05$, 
from here forward we analyze data for $N=5000$ 
only up to depths of $\chi = 0.1$, for which about 10-15 links are present in the boundary layer and the maxima have at least reached the same order of magnitude as the presumed continuum limit. For these cases, the width of the boundary layer does not change significantly when increasing $N$ from 3000 to 5000.  The acceleration in particular is still rising noticeably at this $N$, but more refined spatial scales require impractical times for very deep catenaries because of the very small temporal scales induced by the stringent error tolerances used (Appendix \ref{sec:timeint}).

\section{Ideal and anomalous evolution}

\begin{figure}
    \centering
    \includegraphics[width=6in]{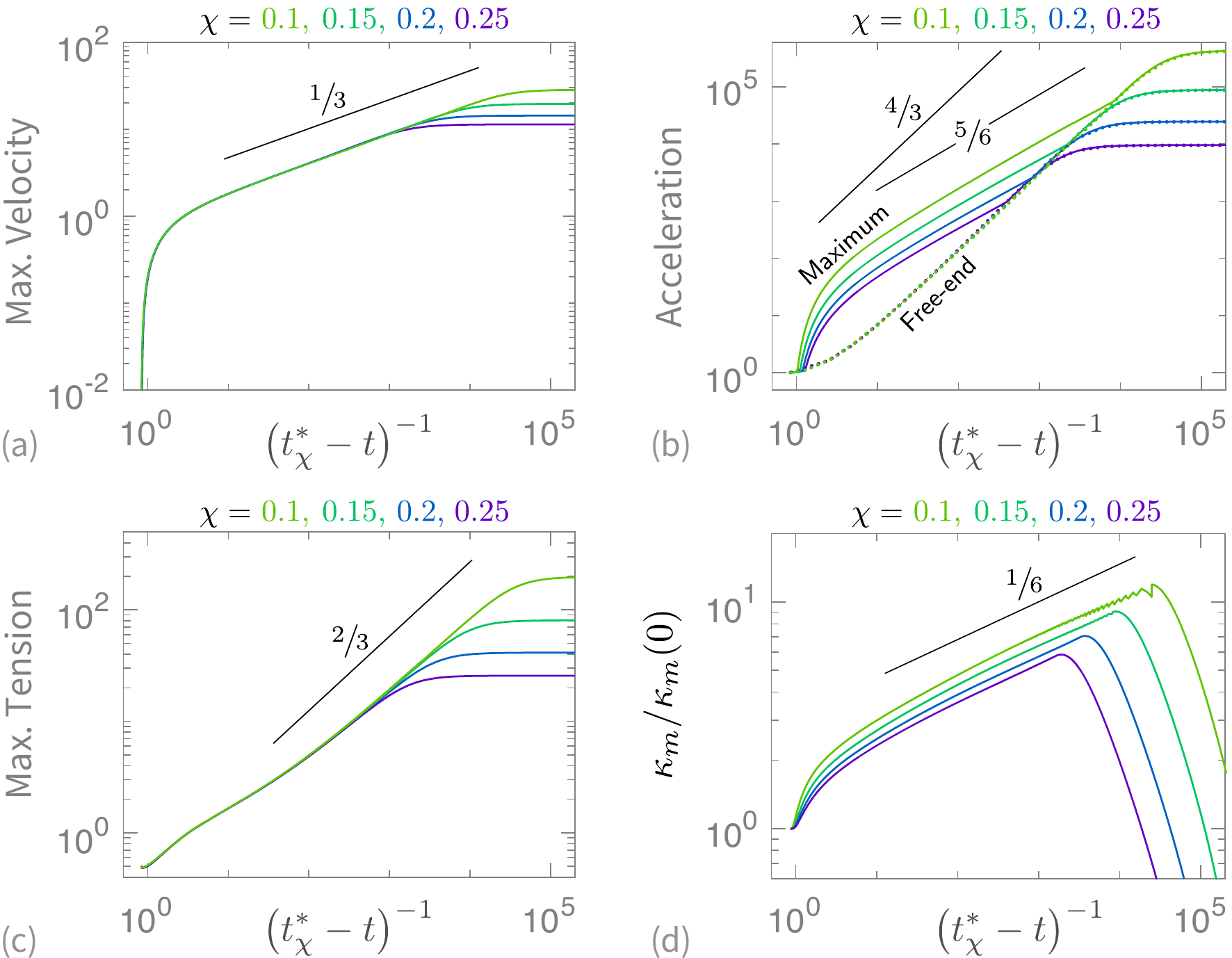}
    \caption{(a)-(c) Evolution of maximum velocity, maximum and free-end accelerations, maximum tension, and (d) amplification of maximum curvature for several catenary depths, using $N=5000$.  Velocity, acceleration, and tension follow the ideal perfectly-folded scalings, while the curvature evolution is anomalously slow.}
    \label{max}
\end{figure}

Given the absence of additional physical effects, we must look to the geometry of the problem for the source of regularization.  
The smallest length scale present in the initial conditions is the initial maximum curvature $\kappa_m(0)$ at the bottom of the hanging catenary, providing a possible scale $\delta_{\kappa_m(0)} = \frac{\pi}{2}\kappa_m^{-1}(0)$ if we envision a whirling of a terminal quarter-circle at the bottom of the string just before snapping.  Quantities derived from this scale are quite large, and just barely provide order-of-magnitude estimates for velocity and tension, but underestimate the acceleration by nearly two orders of magnitude. 
To explain snapping maxima, it is necessary to observe the evolution of quantities, particularly the maximum curvature, as the system approaches the snapping event.  
For much of our data up to the point of regularization, the solution of the singular perfectly-folded string is an excellent guide, even for only moderately deep catenaries. 
Normalizing the analytical results of \cite{Heywood55, CalkinMarch89, OReillyVaradi99, McMillen2005, Virga15, SinghHanna17}, the position of the kink $s_0$ follows the solution of $\ddot s_0 = \tfrac{1}{2}\left( 1 - \tfrac{\dot s_0^2}{1-s_0}\right)$ with initial conditions $s_0 (0) = 0.5$ and $\dot s_0 (0) = 0$.
This may be integrated to obtain 
\begin{align}
    \dot{s}_0 = \pm \dfrac{1}{2}\sqrt{\dfrac{1 - 4(1 - s_0)^2}{2 (1 - s_0)}} \, , \label{shockvelocity}
\end{align}
and the critical snapping time $t^*$, 
\begin{align}
    t^* %= \int\limits_0^{t^*} d t 
    = \int\limits_{0.5}^{1}\! 2\sqrt{\dfrac{2 (1 - s_0)}{1 - 4(1 - s_0)^2}} \, d s_0 
    = \int\limits_0^\frac{1}{\sqrt{2}}\! 2\sqrt{\dfrac{ 1 - 2\zeta^2 }{ 1 - \zeta^2 } } \, d\zeta
    = 2 E\left( \tfrac{\pi}{4} | 2\right) \approx 1.19814 \, , %0235 
\end{align}
using the change of variable $ \zeta^2 = s_0 - \frac{1}{2}$ and the definition of the %incomplete
 elliptic integral of the second kind  \cite{milnethomsonelliptic}. 
%$$ E(a | b) = \int\limits_0^{\sin a} \sqrt{\dfrac{ 1 - b\zeta^2 }{ 1 - \zeta^2 }}  d\zeta$$
The motion of any computed curvature maximum follows this path quite closely before regularization begins. 
The corresponding tether point tension is $\sigma(0,t) = s_0 + \dot s_0^2$.  
As $s_0 \rightarrow 1$, we have $\dot s_0 \sim (1-s_0)^{-\tfrac{1}{2}}$ and $\sigma (1-s_0, t) \sim \sigma(0, t) \sim \dot s_0^2$, 
so the asymptotic scalings of velocity, acceleration, and tension are
\begin{align} \label{asymptoticscalings}
	v \sim \dot s_0 \sim \left(t^* - t\right)^{-\tfrac{1}{3}} \, , \;
	a \sim \ddot s_0 \sim \left(t^* - t\right)^{-\tfrac{4}{3}} \, , \;
	\sigma \sim \left(t^* - t\right)^{-\tfrac{2}{3}} \, .
\end{align}
With these scalings, and individual values of snapping times $t^*_\chi$ observed for each initial condition, the maxima of velocity and tension, as well as the acceleration of the free end, collapse beautifully before peeling off at late times to spread out and approach their regularized values (Fig.\ \ref{max}(a)-(c)). 
At intermediate times, the maximum acceleration (Fig.\ \ref{max}(b)) is located around the fold and follows what appears to be a $(t^*-t)^{-\tfrac{5}{6}}$ scaling, before crossing over to the free end, which follows the ideal scaling. 
 For the catenaries considered in the present study, the values of $t^*_\chi$ obtained from the velocity, acceleration, and tension maxima differ only at the fifth significant digit and so can be thought of as a single value, though for shallow catenaries they may differ considerably.
 
Unfortunately the singular solution cannot directly tell us about curvature, which begins and remains a delta function in those dynamics, but we can infer a scaling.  From dimensional analysis or general considerations about the evolution of curvature \cite{Brower84, GoldsteinPetrich91, Nakayama92},  
we expect $\dot \kappa \sim v \kappa^2$, which along with \eqref{asymptoticscalings} implies a scaling like that of the tension, $\kappa \sim (t^*-t)^{-\tfrac{2}{3}}$. 
Curiously, this corresponds to the observed behavior of Brun and co-workers' integral curvature measure in the boundary-driven unwrapping geometry \cite{Brun16}.
However, in our problem the maximum curvature is observed empirically to evolve according to an anomalously weak scaling, apparently approaching $(t^*-t)^{-\tfrac{1}{6}}$ at important late times preceding regularization (Fig.\ \ref{max}(d)), using values of $t^*_\chi$ obtained from the other quantities.  
In practical terms, the amplification of maximum curvature is weak, which is also in keeping with the observations of \cite{Brun16}, and mostly occurs in a small time window later in the process.  
The maximum reaches the end of the string before rapidly decreasing during the whirling motion preceding the snapping time.
%Finally, as the end whirls around, the maximum curvature falls off rapidly. 
It is plausible that the intermediate-time behavior is linked to that of the maximum acceleration, also unexplained, through the relation between the normal acceleration and the product of tension and curvature.  We note that neither set of curves collapses properly in this regime. 
\begin{figure}
    \centering
    \includegraphics[width=6in]{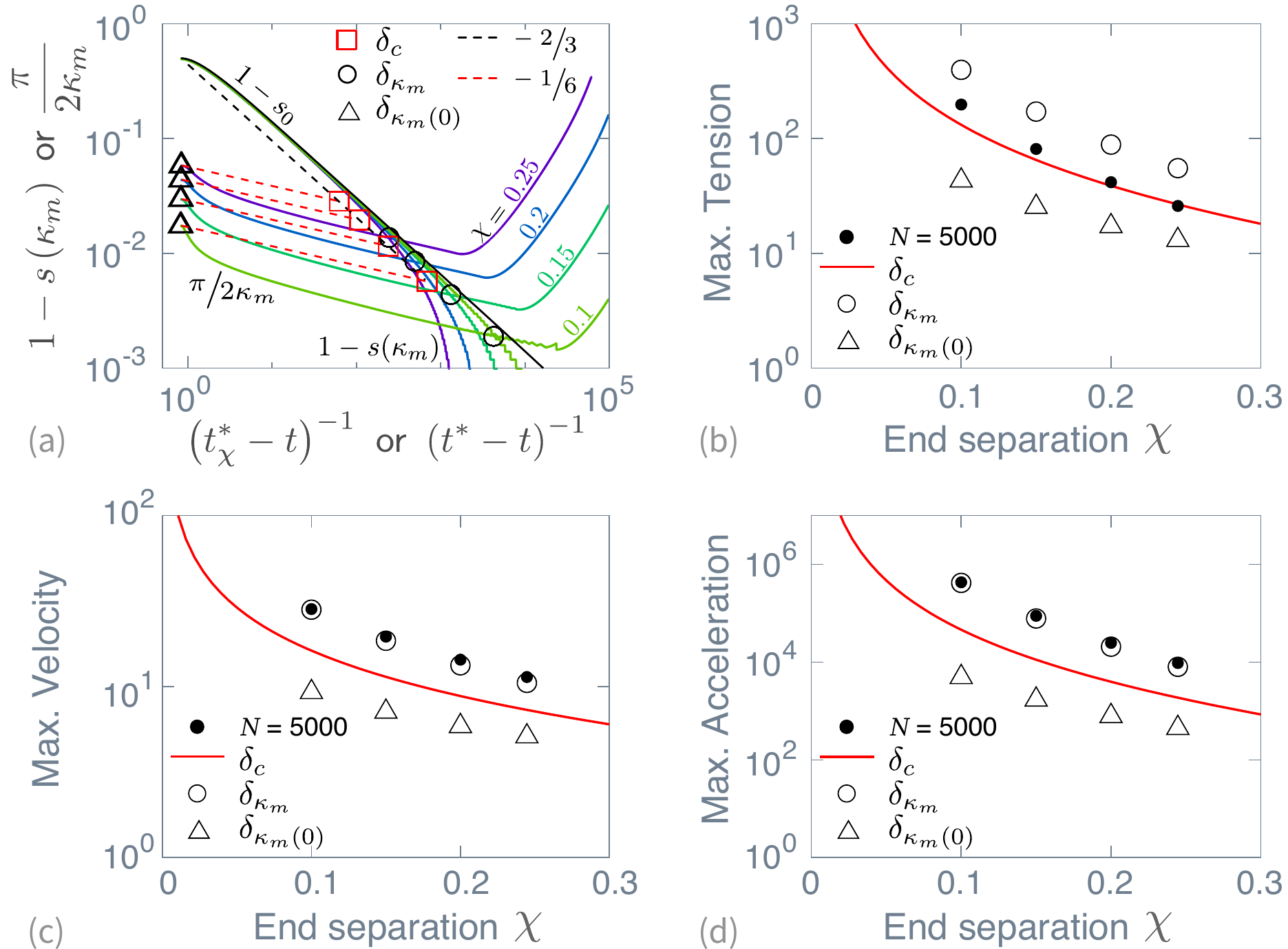}
    \caption{(a) Location of the curvature peak $s(\kappa_m)$ (shown only before it reaches the end of the string), and evolution of the length scale $\frac{\pi}{2}\kappa_m^{-1}$ derived from the curvature peak, for several catenary depths, in numerics using $N=5000$ (solid colored curves) and in linear approximations (dashed lines).  Also shown is the location of the kink $s_0$ in the ideal case (solid black curve). Three possible derived regularizing length scales $\delta$ are shown. 
     (b)-(d)  
Computed maxima of tension, velocity, and acceleration (black dots), and   
predictions obtained through relations \eqref{eq:estimates} for the three scales $\delta$.  Values of $\delta_c$ can be obtained analytically from \eqref{deltacrude}, so those predictions are shown as a continuous curve. }
    \label{theory}
\end{figure}

\section{Amplification estimates} 

This information suggests two possible ways to improve the estimate of the regularization length scale over the $\delta_{\kappa_m(0)}$ based on initial maximum curvature.  
It is reasonable to assume that the transition to whirling behavior of the free end occurs when the distance of the front (associated with the fold in the string, for which the location of maximum curvature $s(\kappa_m)$ is a suitable proxy until regularization) 
from the free end is subtended by the quarter-circle provided by the maximum curvature itself, $1-s(\kappa_m) =\frac{\pi}{2}\kappa_m^{-1}$. 
First, we can numerically compute when these two length scales collide at some $\delta_{\kappa_m}$ (Fig.\ \ref{theory}(a), solid colored curves intersecting in black circles).  Using this scale, the maximum velocity and acceleration obtained through relations \eqref{eq:estimates} are in excellent agreement (Fig.\ \ref{theory}(c)-(d)), although this must be partly fortuitous, as quantities for finite $N$, particularly acceleration, are underestimates of the continuum limit. Meanwhile, the tension is overestimated by about a factor of two for all cases  (Fig.\ \ref{theory}(b)), likely indicating that the steady whirling rod description of snapping needs refinement.  
Alternately, we can provide a cruder estimate of the scale that requires only some knowledge of the initial conditions, the ideal singular solution, and the observed time scaling of curvature.  
This is in the form of the intersection of two straight lines in a log-log plot, starting at the initial conditions but following the late-time scalings (Fig.\ \ref{theory}(a), dashed lines intersecting in red squares).  That is, the distance from the end is approximated as 
$1 - s(\kappa_m) \approx \tfrac{1}{2} \left( \frac{t^* - t}{t^*} \right)^{\tfrac{2}{3}}$ and the length scale as 
$\frac{\pi}{2}\kappa_m^{-1} \approx \frac{\pi}{2}\kappa_m^{-1}(0) \left( \frac{t^* - t}{t^*}\right)^{\tfrac{1}{6}}$, which coincide when $\frac{t^*-t}{t^*} = \left(\frac{\pi}{\kappa_m(0)}\right)^2$ at a scale
\begin{align}
	\delta_c = \frac{1}{2}\left(\frac{\pi}{\kappa_m(0)}\right)^{\tfrac{4}{3}}\,, \label{deltacrude}
\end{align}
using the theoretical ideal $t^*$ for the perfectly-folded string. 
This estimate fails to account for the early-time dynamics of the front and the early and intermediate-time dynamics, including an approximate doubling, of the curvature. It provides a less accurate estimate (Fig.\ \ref{theory}(b)-(d), red curves), but one that is still within the order of magnitude, and practical as it requires neither numerical integration of the full Eqs.\ (\ref{eq:eom}-\ref{eq:eom2}) nor even inversion of the implicit integration of the ideal case \eqref{shockvelocity}.

\section{Concluding remarks} 

In conclusion, we have identified a mechanism for regularization of snapping and provided reasonable practical estimates for the amplification of velocity, acceleration, and tension.
This offers the potential to develop design rules for flexible tethers, stretching energy harvesters, or energy focusing applications.  
Comparison with limited observations on other loading conditions and geometries %(unwrapping, end pulling)
suggests a potential universality of response. 
Further work should explore whether our observed anomalously slow maximum curvature scaling is universal and if so, seek its origin.
It is curious that in the geometrically regularized problem, where finite initial curvature provides the regularization, the evolution of curvature itself also suggests a weak singular scaling. 
%although this is not as certain as the clear collapse of the velocity, acceleration, and tension onto the ideal folded singular scalings.
So although it appears that the physical singularities in the falling perfectly-folded string are a consequence of the singularity in its initial geometry, it may be possible to induce blowup of physical quantities under other, smoother conditions. 
In fact, the expected scaling of inverse maximum curvature would have had the same exponent, $\frac{2}{3}$, as that of the shrinking extent of the boundary layer, which might have precluded the intersection of curves as in Fig.\ \ref{theory}(a) and therefore any geometric regularization of the singularity.

\begin{acknowledgments}
We thank B. Audoly and J.-C. G{\'{e}}minard for sharing data, E. Hamm, A. G. Nair, B. Rallabandi, H. Singh, and E. Vouga for conversations, and C. Rycroft for helpful skepticism and feedback on an early presentation of these results.
\end{acknowledgments}

\appendix
 
\section{Numerical method} \label{sec:timeint}

We employ an adaptation of Preston's finite-difference scheme \cite{preston2011} %(to add gravity and different b.c.)  
along with the Euler-Richardson method \cite{gatland1994}.
The string is discretized into $N+1$ nodes 
connected by $N$ massless links $l_k$, $k=1,\ldots,N$, 
%$(l_1,l_2,\ldots,l_N)$ 
 of equal length $\ell/N$,  
 with a massless node $N+1$ serving as the fixed end (note that these numbers go in the opposite direction as the arc length in our continuous formulation, but the resulting expressions are not affected by the sign changes). The link $l_k$ between the nodes $k$ and $k+1$ bears a tension $\sigma_k$. 
A tensionless ghost-link $l_0$ and massless ghost node are attached to the free end. 
All other nodes are of equal mass.
Adapting  \cite{preston2011}, discrete forms of Eqs.\ (\ref{eq:eom}-\ref{eq:bcs}) are
\begin{align} 
  &(N/\ell)^2 \left\lbrace \sigma_k \left( \bm{r}_{k+1} - \bm{r}_{k} \right) - \sigma_{k-1} \left( \bm{r}_{k} - \bm{r}_{k-1} \right) \right\rbrace +  \rho g\bm{\hat{e}}_2 = \rho d^2_t  \bm{r}_k \, , \label{eq:discretemo} \\
  &(N/\ell)^2 \sigma_{k+1} \left( \bm{r}_{k+2} - \bm{r}_{k+1} \right)\cdot\left( \bm{r}_{k+1} - \bm{r}_{k} \right) - 2\sigma_k 
    + (N/\ell)^2 \sigma_{k-1} \left( \bm{r}_{k+1} - \bm{r}_{k} \right)\cdot\left( \bm{r}_{k} - \bm{r}_{k-1} \right) = - \left| d_t\bm{r}_{k+1} - d_t\bm{r}_{k} \right|^2 \, ,  \\
	&\bm{r}_{N+1} = \bm{0}\, , \quad \sigma_{0} = 0\, , \quad \left(\sigma_{N+1} - \sigma_N \right) + \left(\bm{r}_{N+1} - \bm{r}_N \right)\cdot\rho g\bm{\hat{e}}_2 = 0 \, ,
\end{align}
where we use $\bm{r}_{N+2} - \bm{r}_{N+1} = \bm{r}_{N+1} - \bm{r}_{N}$.
Formulae for angle and curvature are 
\begin{align} 
	\theta_k = \tan^{-1} \dfrac{\left( \bm{r}_{k+1} - \bm{r}_k \right) \cdot \bm{\hat{e}}_2 }{ \left( \bm{r}_{k+1} - \bm{r}_k \right) \cdot \bm{\hat{e}}_1 }  \, , \quad 
	\kappa_k = (N/\ell) \left( \theta_{k+1} - \theta_k \right) \, .
 \end{align}
The discretized momentum equation \eqref{eq:discretemo} could also arise as the Euler-Lagrange equation for an N-pendulum system with discrete Lagrangian
\begin{align}
%L_N = 
&\sum\limits_{k=1}^{N}\left[ (\rho \ell/2N) d_t \bm{r}_k \cdot d_t \bm{r}_k - (\sigma_k /2) \times \left( \left| \bm{r}_{k+1} - \bm{r}_k \right|^2 - \left(\ell/N\right)^2 \, \right) - (\rho g\ell/N)  \left( \ell/2 - \bm{r}_k \cdot \bm{\hat{e}}_2 \right ) \right] \, .
\end{align}
 
The discrete equations for the position and tension are integrated in time as follows.
Given $\bm{r}_k(t)$ and $d_t\bm{r}_{k}(t)$, compute $d_t^2\bm{r}_{k}(t)$ and $\sigma_k(t)$ from the governing equations. 
Then compute the half-step values
\begin{align*}
    \bm{r}_k \left(t + {\Delta t}/{2} \right) &= \bm{r}_k (t) + d_t\bm{r}_{k}(t)\, {\Delta t}/{2}\, ,  \\
    d_t\bm{r}_{k} \left(t + {\Delta t}/{2} \right) &= d_t\bm{r}_{k} (t) + d_t^2\bm{r}_{k}(t)\, {\Delta t}/{2}\, ,
\end{align*}
and use these to compute $\sigma(t + \Delta t/2)$ and $d_t^2\bm{r}_{k} (t + {\Delta t}/{2} )$ from the governing equations.
Then compute the full-step values
\begin{align*}
    \bm{r}_k \left(t + {\Delta t} \right) &= \bm{r}_k (t) + d_t\bm{r}_{k}(t+{\Delta t}/{2})\, {\Delta t}\, , \\
    d_t\bm{r}_{k} \left(t + {\Delta t} \right) &= d_t\bm{r}_{k} (t) + d_t^2\bm{r}_{k}(t+{\Delta t}/{2})\, {\Delta t}\, ,
\end{align*}
and use these to compute $\sigma(t + \Delta t)$ and $d_t^2\bm{r}_{k} (t + {\Delta t} )$ from the governing equations.
This formulation is accurate to order $(\Delta t)^3$ in positions and velocities, with
an error estimate at each time step of \cite{gatland1994}
\begin{align}
\varepsilon = ({\Delta t}/{2}) \max \left( \big| d_t\bm{r}_{k}\left(t + {\Delta t}/{2} \right) - d_t\bm{r}_{k}\left(t \right) \big| , 
\big| d_t^2\bm{r}_{k}\left(t + {\Delta t}/{2} \right) - d_t^2\bm{r}_{k}\left(t \right) \big| \right)  \, .
\label{eq:error}
\end{align} 
To keep the computations within a desired tolerance $\varepsilon_\text{tol} = 10^{-8}$, the time step is adapted as follows \cite{gatland1994}:
\begin{equation}
    \Delta t_\text{new} = 0.9 \Delta t_\text{old} ({\varepsilon_\text{tol}}/{\varepsilon})^{1/2} \, .
    \label{eq:dtau}
\end{equation}
Physical measures of error are any stretching of the string or change in energy.  
Changes in normalized local link length, global string length, and global energy are kept below
$5 \times 10^{-4}$, $4 \times 10^{-6}$, and $8 \times 10^{-7}$, respectively, for the computations reported. 
The largest errors do not occur during the snapping events; the largest change in normalized local link length during snapping is two orders of magnitude smaller.  
%changes in normalized local link length and global energy are kept below $3 x 10^{-6}$ and $3 x 10^{-8}$, respectively, during snapping. 
 This stringent tolerance results in smallest time steps 4-5 orders of magnitude smaller than the $\chi$-dependent snapping event durations of $10^{-3}$ to $10^{-5}$.  
 %, $10^{-7}$ to $10^{-10}$ depending on $\chi$. 
Adjusting the tolerance by two orders of magnitude from $10^{-7}$ to $10^{-9}$ changes the time step by an order of magnitude but has negligible effects on the maximum values of interest; with $N=5000$ the maxima change at the ninth significant digit for velocity and the sixth significant digit for acceleration and tension.

%\bibliography{refs_snap.bib} 
\bibliographystyle{unsrt}

\end{document}